\documentclass{iopart}
\usepackage{graphicx}
\usepackage{graphicx}   
\usepackage{mathrsfs}
\usepackage{color}

\begin{document}

\title{Graphene transparency in weak magnetic fields}

\author{David Valenzuela$^1$, Sa\'ul Hern\'andez-Ortiz$^2$, Marcelo Loewe$^{1,3}$, Alfredo Raya$^{1,2}$}

\address{$^1$Instituto de F\'{\i}sica, Pontificia Universidad Cat\'olica de Chile, 
Casilla 306, Santiago 22, Chile.\\
$^2$Instituto de F\'{\i}sica y Matem\'aticas, Universidad Michoacana de San Nicol\'as de Hidalgo, 
Edificio C-3, Ciudad Universitaria, 58040 Morelia, Michoac\'an, M\'exico.\\
$^3$Centre for Theoretical and Mathematical Physics and Department of 
Physics, University of Cape Town, 
Rondebosch 7700, South Africa. \\
devalenz@uc.cl, sortiz@ifm.umich.mx, raya@ifm.umich.mx, mloewe@fis.puc.cl}

\begin{abstract}
We carry out an explicit calculation of the vacuum polarization tensor for an effective low-energy model of monolayer graphene in the presence of a weak magnetic field of intensity $B$ perpendicularly aligned to the membrane. By expanding the quasiparticle propagator in the Schwinger proper time representation  up to order $(eB)^2$, where $e$ is the unit charge, we find an explicitly transverse tensor, consistent with gauge invariance. Furthermore, assuming that graphene is radiated with monochromatic light of frequency $\omega$ along the external field direction, from the modified Maxwell's equations we derive the intensity of transmitted light and the angle of polarization rotation in terms of the longitudinal ($\sigma_{xx}$) and transverse ($\sigma_{xy}$) conductivities. Corrections to these quantities, both calculated and measured, are of order $(eB)^2/\omega^4$. Our findings generalize and complement previously known results reported in literature regarding the light absorption problem in graphene from the experimental and theoretical points of view, with and without external  magnetic fields.
\end{abstract}

\submitto{\JPA}
\maketitle
\section{Introduction}	

One decade has gone by since the earlier groundbreaking experiments performed by Andrei Geim and Konstantin Novoselov~\cite{graphene1} (Nobel Laureates in Physics in 2010) to isolate single layer membranes of graphite, graphene. Soon after, theoretical~\cite{graphene2} and experimental~\cite{graphene3} groups highlighted the properties of charge carriers in this material which resemble much to ultrarelativistic electrons, thus establishing a bridge between solid state and particle physics (see, for instance, Refs.~\cite{Gusynin:review,review2}). Graphene has given rise to the new era of Dirac materials with potential applications in nanotechnology, but also offering an opportunity to test the core of fundamental physics in a condensed matter environment. Mechanical, thermal and electronic properties of this two-dimensional crystal locate it among the best candidates to replace silicon in nanotechnological devices, basically due to its hardness, yet flexibility, high electron mobility and thermal conductivity~\cite{properties}. 

Crystal structure of graphene consists in a honeycomb array of tightly packed carbon atoms, thus allowing an accurate tight-binding description. At low energies, such a description becomes in the continuous limit the Lagrangian of massless quantum electrodynamics in (2+1)-dimensions, QED$_3$, for the charge carriers restricted to move along the membrane~\cite{Gusynin:review}, but in which the ``photon'' is allowed to move throughout space in such a way that the static Coulomb interaction is still described by a potential that varies as the inverse of the distance on the plane of motion of electrons. In this form, low-energy dynamics of graphene is in accordance with the spirit of brane-world scenarios of fundamental interactions (see, for instance, Ref.~\cite{RQED}) where the gauge field (photon) is allowed to move throughout the bulk (full space), but matter fields are restricted to a brane (the graphene layer).

Expectedly, quantum field theoretical methods have been developed to describe phenomena in graphene which have been theorized in the high energy physics realm, but that would appear enhanced in this material due to the ratio of the speed of light in vacuum and the Fermi velocity of its charge carriers, $c/v_F\simeq 300$. Theoretical objects like the effective action  in external electromagnetic fields have been calculated by several authors in connection with the Schwinger mechanism for pair production and the issue of minimal conductivity~\cite{effact}, ideas that have been generalized to the multilayer case~\cite{Katsnelson}. Other ``relativistic'' effects discussed in literature  include the Klein paradox~\cite{klein}, Casimir effect~\cite{casimir} and the dynamical formation of a mass gap from excitonic condensates~\cite{DMG}. Graphene properties have been handled also from the perspective of non-conmutative quantum mechanics~\cite{NCQM}.

A remarkable feature of graphene is the visual transparency of the membranes. Its opacity has been measured~\cite{measure} to be roughly 2.3\% with almost negligible reflectance. This observation has opened the possibility of using single layers of this crystal in combination with bio-materials to produce clean hydrogen by photocatalysis~\cite{hydrogen} with visible light. The problem of light absorption in graphene can be addressed from quantum field theoretical methods~\cite{fial1}. Several authors have considered  the Dirac picture for its charge carriers  in terms of the degrees of freedom of  QED$_3$ under different assumptions.  Parity violating effects were considered in~\cite{fial2}, whereas the influence of a strong magnetic field was considered in~\cite{fial3} in connection with the Faraday effect. Measurements of magneto-optical properties of epitaxial graphene have been reported in Ref.~\cite{faradayexp}, in particular the polarization rotation and light absorption. Results seem to be in accordance with the ``relativistic'' behavior of charge carriers for a range of values of the external magnetic field intensity between 0.5~-~7~T~\cite{fial3}. For the discussion of these results, the structure of the vacuum polarization tensor is the cornerstone. This operator has been calculated by several authors in the presence of a strong magnetic field perpendicularly aligned with the graphene membrane~\cite{strongB}. In this work, we continue the discussion but in our considerations, the external magnetic field is weak in intensity as compared to the effective mass $\Delta^2=(p_F/v_F)^2$, where $p_F$ and $v_F$ are, respectively, the Fermi momentum and Fermi velocity of charge carriers. 
The article is organized as follows: We start modeling the low-energy behavior of graphene from massless QED$_3$ subjected to an external magnetic field perpendicular to the membrane, namely, we consider the full space, but restrict the dynamics of charge carriers in graphene to an infinite plane where the third spatial component is set to zero. Expanding the quasiparticle propagator in the weak field regime, we calculate the vacuum polarization tensor to the leading order in the external field intensity in Sect. 2. In Sect.~3, we introduce the polarization operator in the modified Maxwell's equation to describe the propagation of electromagnetic waves in space. From the matching conditions, we calculate the transmission coefficient and from there, the intensity of transmitted light and angle of polarization rotation in terms of the longitudinal and transverse conductivities,  which we derive from Ohm's law.  Our results correspond to the weak field Faraday effect. We discuss our findings and conclude in Sect.~4. Some details of the calculation of the polarization tensor are presented in an appendix.

\section{A continuous model for graphene}

Tight-binding approach to the description of monolayer graphene corresponds in the continuum to a massless version of quantum electrodynamics in (2+1) dimensions, but with a static Coulomb interaction which varies as the inverse of the distance, just as in ordinary space~\cite{Gusynin:review}. We adopt the conventions of Refs.~\cite{fial1,fial2,fial3} and consider an infinite graphene membrane immersed in a (3+1)-dimensional space oriented along the plane $z=0$. The action for this model is expressed as
\begin{equation}
S=-\frac{1}{4}\int d^4x F_{\mu\nu}^2+\int d^3x \bar\psi {\not \! D}\psi\;,
\label{action}
\end{equation}
with $F_{\mu\nu}=\partial_\mu A_\nu-\partial_\nu A_\mu$ and ${\not \! D}=i\tilde{\gamma}^a(\partial_a+ieA_a)$. In our considerations, greek indices take the values 0,1,2,3, and latin indices 0,1,2, labeling the coordinates of the graphene layer. Moreover, the re-scaled Dirac matrices are such that $\tilde{\gamma}^0=\gamma^0$, $\tilde{\gamma}^{1,2}=v_F\gamma^{1,2}$ and for later convenience, we also consider the matrix $\tilde{\gamma}^3=\gamma^3$, where $v_F$ is the Fermi velocity of quasiparticles in the crystal. In the natural units of the system (namely, when $v_F=1$), the form of the action has been dubbed as Reduced QED and has been proposed in the context of brane-world scenarios~\cite{RQED}.

\begin{figure}[t!]
\begin{center}
\includegraphics[width=0.5\textwidth]{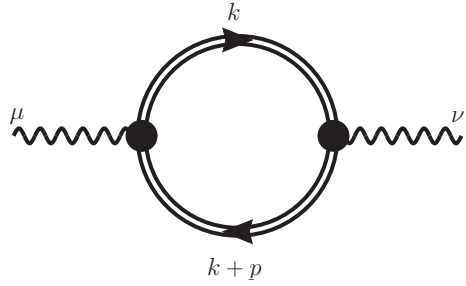}
\end{center}
\caption{Vacuum polarization diagram.}
\label{fig1}
\end{figure}

Measuring the response of graphene to external electromagnetic fields amounts to calculate the effective action, which in turn is expressed through the vacuum polarization tensor $\Pi^{\mu\nu}$. Because in this case the dynamics of fermions is restricted to a plane 
according to Fig.~\ref{fig1} we can express 
\begin{equation}
\Pi^{\mu \nu}(p)=ie^2{\rm Tr}\Bigg[\int_{-\infty}^\infty dk_z\delta(k_z)\int \frac{d^3k}{(2\pi)^3}\tilde{\gamma}^\mu S(k) \tilde{\gamma}^\nu S(k+p)\Bigg]\;,\label{fullvp}
\end{equation}
where the trace is over full space and then we set $\Pi^{\mu 3}=\Pi^{3 \mu}=0.$ 
 Here, $S(p)$ represents the quasiparticle propagator (electric charge $-e$) and the double fermion line in the diagram specifies that the propagator is corrected by some classical external field. We consider the situation in which a uniform magnetic field is aligned perpendicularly to the graphene membrane. We think of this field as being weak in intensity, as compared to the the natural scale $\Delta^2=(p_F/v_F)^2$, where $p_F$ is the quasiparticles Fermi momentum such that $\Delta$ behaves as an effective Dirac mass for the charge carriers. This situation can be formally achieved by considering the quasiparticles with a finite mass gap $\Delta$  and then expand the corresponding Schwinger propagator in the proper time representation~\cite{Schwinger},
\begin{eqnarray}
iS(p)&=& \int_0^\infty ds\, e^{is\left(p_\parallel^2+p_\perp^2 \frac{\tan{(eBs)}}{eBs}-\Delta^2+i\epsilon \right)}\nonumber\\
&&\hspace{-7mm}
\times \left[ (\tilde{\gamma}\cdot p_\parallel+\Delta)(1+\gamma^1\gamma^2 \tan{(eBs)})+\tilde{\gamma}\cdotp_\perp(1+\tan^2{(eBs)})\right]\,,\label{PTS}
\end{eqnarray}
in powers of $(eB/\Delta^2)$, retaining terms up to order ${\cal O}((eB/\Delta^2)^2)$ and then letting $\Delta\to 0$~\footnote{We emphasize that the Schwinger phase that accompanies the fermion propagator (\ref{PTS}) in the proper time representation does not contribute in the vacuum polarization tensor, and thus we neglect it from start.}.  We adopt a prescription where we split the transverse and parallel components  --with respect to the magnetic field direction-- of an arbitrary vector $v^a$ defined on the graphene membrane according to $v^a=(v_\parallel,v_\perp)$ such that $v^2\geq 0$. Any reference to the third spatial component has been taken into account in the $\delta(k_z)$ integration in Eq.~(\ref{fullvp}) and does not appear in what follows. Therefore, $\tilde{\gamma}\cdot v=\tilde{\gamma}\cdot v_\parallel+\tilde{\gamma}\cdot v_\perp$ and $v^2=v_\parallel^2+v_\perp^2$. Furthermore, we take $g^{a b}={\rm diag}(1,-1,-1)\equiv g^{a b}_\parallel+g^{a b}_\perp$, such that $g^{a b}_\parallel={\rm diag}(1,0,0)$. Thus, in the weak field limit,  the  structure of the quasiparticle propagator becomes~\cite{taiwaneses}
\begin{eqnarray}
S(p)&=&S_0(p)+eB S_1(p)+(eB)^2 S_2(p)\nonumber\\
&\equiv& \frac{(\tilde{\gamma}\cdot p)}{p^2}+i eB \frac{\tilde{\gamma}\cdot p_\parallel\gamma^1\gamma^2}{(p^2)^2} +\frac{2(eB)^2}{(p^2)^4}\left[p_\perp^2 \tilde{\gamma}\cdot p_\parallel-p_\parallel^2 \tilde{\gamma}\cdot p_\perp \right]\,. 
\label{expansionS}
\end{eqnarray}
Here, the matrices $\gamma^1$ and $\gamma^2$ do not appear rescaled because the operators ${\cal O}^\pm=(I\pm \gamma^1\gamma^2)/2$, with $I$ the identity matrix, correspond to the (pseudo)spin projection operators~\cite{taiwaneses}.
With the above expansion~(\ref{expansionS}), it is straightforward to verify that the structure of the vacuum polarization is
\begin{equation}
\Pi^{\mu \nu}(p)= \eta^\mu_a \left[ \Pi^{a b}_{(0)}(p)+ (eB)^2 \Pi^{a b}_{(2)}(p)\right] \eta_b^\nu\;,\label{fullPi}
\end{equation}
where we have defined $\eta^\mu_a={\rm diag}(1,v_F,v_F)$. The first term in the square bracket represents the polarization tensor in vacuum, whereas the second term stands for the  quadratic order contribution to the polarization tensor. The linear correction in $(eB)$, $\Pi_{(1)}^{a b}(p)$, is absent due to the parity preserving property of the model. In other words, contributions to the polarization arising from a Chern-Simons term are not considered in this work. 

The  magnetic field independent vacuum polarization tensor $\Pi_{(0)}^{a b}$ 
has been calculated by many authors~\cite{fial1,vacresult}. It is of the form
\begin{equation}
\Pi_{(0)}^{a b} = 4\pi \tilde{\alpha} \Pi_{\rm vac}(p) \left(g^{a b}- \frac{\tilde{p}^a \tilde{p}^b}{\tilde{p}^2} \right)\;,
\end{equation}
with $\tilde{\alpha}=\alpha/v_F^2$ and $\alpha=e^2/(4\pi)$ as usual. Moreover, $\tilde{p}$ is the magnitude of the momentum vector with components $\tilde{p}^m=\eta^m_n p^n$, and the polarization scalar
\begin{equation}
\Pi_{\rm vac}(p)=\frac{i}{8} \tilde{p}\;.
\label{vacuum}
\end{equation}
This vacuum contribution is transverse, as demanded by gauge invariance.

On the other hand, the quadratic correction has two contributions, 
\begin{eqnarray}
\Pi^{a b}_{(2)}&=&\Pi^{a b}_{(2)-11}+2 \Pi^{a b}_{(2)-20}\nonumber\\
&=& \int \frac{d^3k}{(2\pi)^3}{\rm Tr}[\tilde{\gamma}^a S_1(k) \tilde{\gamma}^b S_1(k+p)] \nonumber\\
&&+ 
2 \int \frac{d^3k}{(2\pi)^3}{\rm Tr}[\tilde{\gamma}^a S_2(k) \tilde{\gamma}^b S_0(k+p)]\;,\label{pi1120}
\end{eqnarray}
with a suggestive notation that the $\Pi^{a b}_{(2)-11}$ contributions comes from each of the quasiparticle propagators being dressed at the first order in the external field, whereas $\Pi^{a b}_{(2)-20}$ has one propagator without field, whereas the second one is dressed at order $(eB)^2$. The factor of 2 is a symmetry factor. Evaluation of these integrals is cumbersome, but straightforward. Our procedure was the following, we have started by
inserting the expansion in Eq.~(\ref{expansionS}) into each of the contributions to the polarization tensor in Eq.~(\ref{pi1120}). Then, with the aid of the identity
\begin{equation}
\frac{1}{A^p B^q}= \frac{\Gamma(p+q)}{\Gamma(p)\Gamma(q)}\int_0^1 dx \frac{x^{p-1}(1-x)^{q-1}}{[A x + B (1-x)]^{p+1}}\;,
\end{equation}
followed by the shift of variables $k\to k-p(1-x)$, after taking the traces over full space and performing the remaining contractions,  we obtain
\begin{eqnarray}
\Pi^{a b}_{(2)-11}&=& \frac{3i \tilde{\alpha}}{\pi^3}g^{a b}\left[ I^{11}_{104}(\tilde{p})-\tilde{p}_\parallel^2I^{22}_{004}(\tilde{p})\right]\;, \nonumber\\
 \Pi^{a b}_{(2)-20}&=&\frac{4i\tilde{\alpha}}{\pi^3}\Bigg[\left(g_\parallel^{ab}-g_\perp^{ab}\right)\left(I^{03}_{115}(\tilde{p}) +\tilde{p}_\perp^2 I^{23}_{105}(\tilde{p})\right)\nonumber\\
&& +g_\parallel^{ab}\left( I^{03}_{115}(\tilde{p})+p_\parallel^2I^{23}_{015}(\tilde{p})\right) \nonumber\\
&&-\left(\tilde{p}_\parallel^a \tilde{p}^b +\tilde{p}_\parallel^b \tilde{p}^a-\tilde{p}_\parallel^2g^{ab}\right)\left(I^{14}_{015}(\tilde{p})+\tilde{p}_\perp^2 I^{23}_{005}(\tilde{p})\right)\nonumber\\
&& + \left(\tilde{p}_\perp^a \tilde{p}^b +\tilde{p}_\perp^b \tilde{p}^a-\tilde{p}_\perp^2g^{ab}\right)
\left(I^{14}_{105}(\tilde{p})+\tilde{p}_\parallel^2 I^{23}_{005}(\tilde{p})\right)\nonumber\\
&&+\left(\tilde{p}_\perp^a \tilde{p}^b_\parallel +\tilde{p}_\parallel^a \tilde{p}^b_\perp\right) \left(I^{23}_{015}(\tilde{p})-2I^{23}_{105}(\tilde{p}) \right)\Bigg]\;,
\end{eqnarray}
where the master integral
\begin{eqnarray}
I^{{\rm fg}}_{{\rm mnr}}(\tilde{p})&=&\int_0^1 x^{\rm f} (1-x)^{\rm g} \int d^3k \frac{(k_0^2)^{\rm m}(k_\perp^2)^{\rm n}}{[k^2+\tilde{p}^2x(1-x)]^{\rm r}}\;,\nonumber\\
&=& (-1)^{\rm m+n-r}\frac{i\pi}{(\tilde{p}^2)^{{\rm r-m-n}-3/2}} B\left({\rm n}+1,{\rm r-n}-1\right)\nonumber\\
&&\hspace{-34mm}\times
B\left({\rm m}+\frac{1}{2},{\rm r-m-n}-\frac{3}{2} \right) B\left(\rm{f-r-m-n}+\frac{5}{2},{\rm g-r-m-n}+\frac{5}{2} \right)\;,
\label{master}
\end{eqnarray}
is written in terms of beta functions $B(x,y)$ and whose explicit evaluation is presented in the appendix.
Making use of the master integral, the quadratic correction in the external field to the polarization tensor can be written as
\begin{equation}
\Pi^{\mu\nu}_{(2)}=4\pi\tilde{\alpha}\eta^\mu_a \left[ \Pi_0(\tilde{p}) {\cal P}^{ab} + \Pi_\perp(\tilde{p}) {\cal P}_\perp^{ab}\right] \eta_b^\nu\,,
\label{Pi2}
\end{equation}
with the transverse tensors
\begin{eqnarray}
{\cal P}^{ab}&=&\left(g^{ab}-\frac{\tilde{p}^a \tilde{p}^b}{\tilde{p}^2} \right)\;,\qquad
{\cal P}^{ab}_\perp\ =\ \left(g^{ab}_\perp-\frac{\tilde{p}^a_\perp \tilde{p}^b_\perp}{\tilde{p}^2_\perp} \right)\;,
\label{projectors}
\end{eqnarray}
and the polarization scalars
\begin{eqnarray}
\Pi_0(\tilde{p})&=&\frac{i}{8\tilde{p}^3}\left(1-5\frac{\tilde{p}_\parallel^2}{\tilde{p}^2} \right)\;,\quad
\Pi_\perp(\tilde{p})\ =\ \frac{i}{4\tilde{p}^3}\left( 1-\frac{\tilde{p}_\parallel^2}{\tilde{p}^2}\right)\;.
\label{scalars}
\end{eqnarray}
Thus, the final expression for $\Pi^{\mu \nu}$ becomes
\begin{eqnarray}
\Pi^{\mu\nu}(p)&=&4\pi\tilde{\alpha}\eta^\mu_a \left[\left(\Pi_{\rm vac}(\tilde{p})+(eB)^2\Pi_0(\tilde{p})\right){\cal P}^{a b}
+(eB)^2 \Pi_\perp(\tilde{p}){\cal P}^{a b}_\perp \right]\eta_b^\nu
\;.\nonumber\\
\label{final}
\end{eqnarray}
The above result, Eq.~(\ref{final}), comprises the main result of this section and is the basis for our discussion below. Before proceeding, a few comments are at hand:
\begin{itemize}
\item $\Pi^{\mu\nu}(p)$ is a transverse tensor order by order in $(eB)$. This fact justifies that our procedure to include the influence of the external magnetic field by means of expansion of the proper time representation of the quasiparticle propagator preserves gauge invariance.
\item Our procedure is an alternative to the traditional approach in which the vacuum polarization tensor is expressed as a double proper time integral~\cite{Schwinger,Dittrich,schubert,Shpagin}. In fact, for the particular case of QED in (2+1)-dimensions considered in Ref.~\cite{Shpagin}, the weak field expansion of the polarization scalars, Eqs.~(48)-(50) of that reference, match our findings in the massless limit, when we set $v_F=1$.
\end{itemize}

We shall use the expressions for $\Pi^{\mu\nu}$ developed in this section to discuss the problem of light absorption in graphene.

\section{Light Absorption}

From the action of our model, Eq.~(\ref{action}), we can describe the propagation of electromagnetic waves throughout space according to the modified Maxwell's equations
\begin{equation}
\partial_\mu F^{\mu\nu}+\delta(z) \Pi^{\nu\rho}A_\rho=0\;,\label{maxwell}
\end{equation}
which fulfill the conditions
\begin{eqnarray}
A_\mu\Bigg|_{z=0^+}-A_\mu\Bigg|_{z=0^-}&=&0\;,\nonumber\\
(\partial_z A_\mu)\Bigg|_{z=0^+}-(\partial_z A_\mu)\Bigg|_{z=0^-}&=&\Pi_\mu^\nu A_\nu \Bigg|_{z=0}\;.
\label{bc}
\end{eqnarray}
Following Refs.~\cite{fial1,fial2,fial3}, we interpret the delta function in Eq.~(\ref{maxwell}) as a current along the graphene plane. Thus, from Ohm's law, 
\begin{equation}
j_a=\sigma_{ab} E_b\;,
\label{ohm}
\end{equation} 
where the indices $a,\ b$ take the values 1 and 2, emphazising that they refer to the spatial coordinates of the graphene membrane.
Assuming a varying electric field with frequency $\omega$ expressed in a temporal gauge $A_0=0$, namely, $E_b=i\omega A_b$ and noticing, from the generalized Maxwell's equations~(\ref{maxwell}) that $j_a\simeq \Pi_{ab} A_b$, we can identify the transverse conductivity as
\begin{equation}
\sigma_{ab}=\frac{\Pi_{ab}}{i\omega}\;.
\label{transversesigma}
\end{equation}  
For the problem of light absorption, let us consider a plane wave of frequency $\omega$, which travels along the $z$-direction from below the graphene layer with a linear polarization along the $\hat{e}_x$ direction. 
These assumptions allow us to write~\cite{fial3}
\begin{equation}
\Pi^{jk}(w)=\left( \begin{array}{cc} \Pi_0(w) & 0 \\ 0 & i\omega(\sigma_{xx}\delta^{ab}+\sigma_{xy}\epsilon^{ab})\end{array}
\right)\;,\label{inverse}
\end{equation}
where $\epsilon^{ab}$ is the Levi-Civita symbol and $\sigma_{xx}$, $\sigma_{xy}$ represent the longitudinal and transverse conductivities.  Moreover, considering that the wave insides on the graphene plane, the reflected and transmitted waves can be described as 
\begin{equation}
A=e^{-i\omega t}\left\{ \begin{array}{cc} \hat{e}_x e^{ik_z z}+(r_{xx}\hat{e}_x+r_{xy}\hat{e}_y)e^{-ik_z z}, & z<0\\
(t_{xx}\hat{e}_x+t_{xy}\hat{e}_y)e^{ik_z z}, & z>0\end{array}
\right.
\end{equation}
where $\hat{e}_{x,y}$ are the unit vectors along the directions $x$ and $y$ on the membrane. Thus, from the general form of the vacuum polarization tensor, Eq.~(\ref{inverse}), the boundary conditions~(\ref{bc}) simplify to
\begin{eqnarray}
A_a\Bigg|_{z=0^+}-A_a\Bigg|_{z=0^-}&=&0\;,\nonumber\\
(\partial_z A_a)\Bigg|_{z=0^+}-(\partial_z A_a)\Bigg|_{z=0^-}&=&\alpha \Psi(\omega) \delta^{ab} A_b \Bigg|_{z=0}\;,
\label{bc2}
\end{eqnarray}
where
\begin{equation}
\Psi(\omega)=\alpha\Bigg[\Pi_{\rm vac}(\omega)+(eB)^2\Pi_0(\omega)\Bigg]\;.
\end{equation}
Thus, the transmission coefficients can be straightforwardly obtained~\cite{fial1,fial2,fial3}
\begin{equation}
t_{xx}=\frac{2\omega}{i\alpha\Psi_N(\omega)+2\omega}\;,\qquad t_{xy}=0\;,
\end{equation}
with $\Psi_N(\omega)=N\Psi(\omega)$, accounting for the degrees of freedom of charge carriers. 
Therefore, the intensity of transmitted light is
\begin{equation}
{\cal I}=|t_{xx}|^2\simeq 1+\frac{\alpha {\rm Im}\Psi_N(\omega)}{\omega}+{\cal O}(\alpha^2)\;.
\end{equation}
In terms of the conductivity tensor $\sigma$, ${\cal I}$ and the angle of polarization rotation can be expressed as
\begin{equation}
\theta=-\frac{{\rm Re}\sigma_{xy}}{2}+ {\cal O}(\alpha^2)\;,\qquad {\cal I}=1-{\rm Re}\sigma_{xx}+{\cal O}(\alpha^2)\;.
\label{faraday}
\end{equation}
Substituting the explicit form of the polarization scalars, we finally arrive at the main results of this article, namely,
\begin{equation}
{\cal I}=1-\alpha \pi\left(1 +4 \frac{ (eB)^2}{\omega^4}\right)\;,\qquad \theta=-2\pi\alpha \frac{(eB)^2}{\omega^4}.
\end{equation}
\begin{figure}[t!]
\begin{center}
\includegraphics[width=0.4\textwidth]{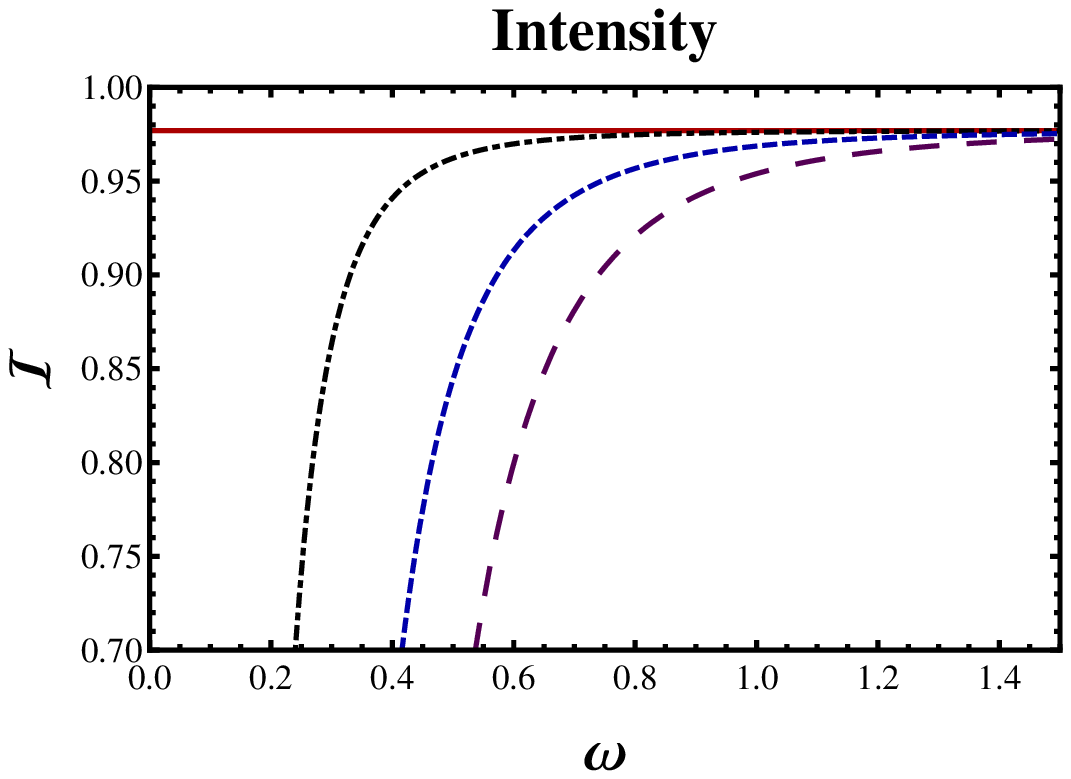}
\includegraphics[width=0.4\textwidth]{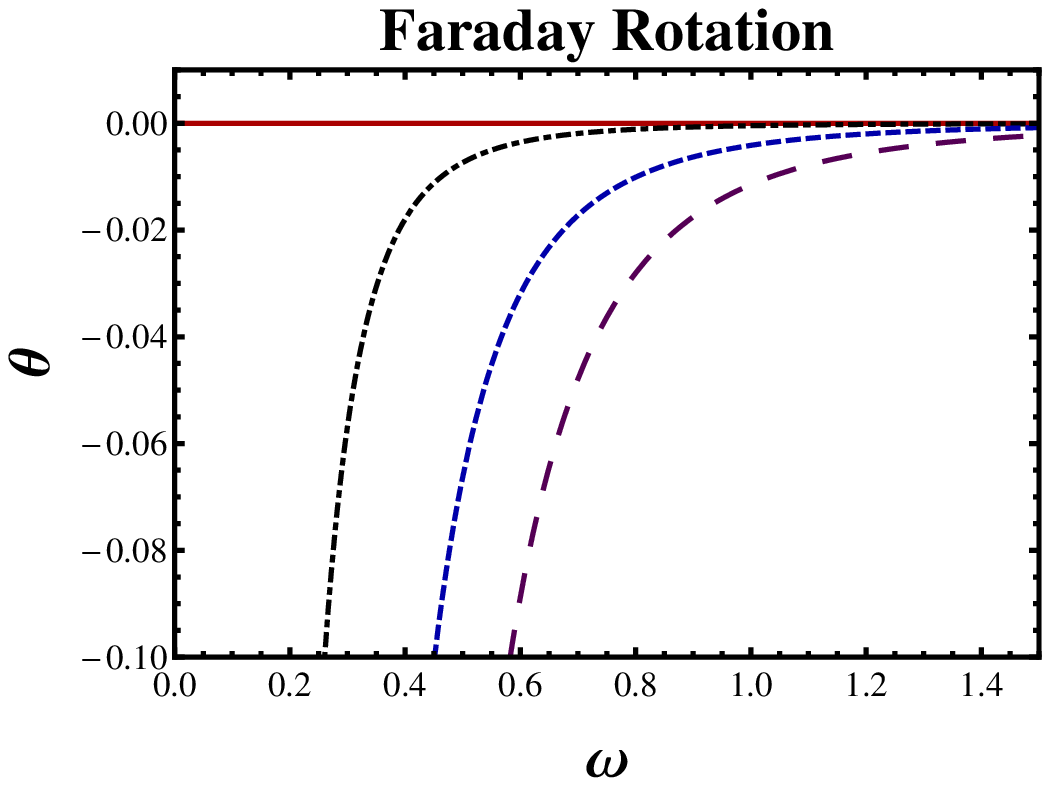}
\end{center}
\caption{Instensity of transmitted light ${\cal I}$ and angle of polarization rotation $\theta$ as a function of the incoming electromagnetic wave frequency $\omega$ (in arbitrary units) for different values of the external magnetic field, also in arbitrary units, but preserving the weakness of the intensity of our approximation. Solid red curve corresponds to the case of $eB=0$ in this set of arbitrary units, dot-dashed black curve, $eB=0.1$, short-dashed blue curve, $eB=0.3$ and long-dashed purple curve, $eB=0.5$.}
\label{fig2}
\end{figure}
These quantities are plotted in the left and right panel, respectively, of Fig.~\ref{fig2} as a function of the frequency of incident light $\omega$ for several values of the external magnetic field.
Comparing with the measured universal absorption rate $\alpha\pi=2.3\%$~\cite{measure}, we conclude that 
in the weak field limit, the intensity of transmitted light and angle of polarization rotation get corrected by factors $(eB)^2/\omega^4$, in consistency
with the experimental and theoretical findings for these quantities in absence of external fields as well as in and the presence of a strong magnetic field~\cite{fial1,faradayexp,strongB}.

\section{Final remarks}

In this work, we have calculated the vacuum polarization tensor in a low energy effective model of graphene based on massless QED$_3$. We have considered a uniform magnetic field aligned perpendicularly to the graphene membrane and expanded the charge carrier propagator in the weak field regime, as compared to the effective mass $\Delta^2=(p_F/v_F)^2$ of the quasiparticles. We have considered the explicit limit $\Delta\to 0$. The Passarino--Veltman-type of integrals involved in the calculation of the polarization operator were obtained after a lengthy, but straightforward procedure from a single master integral that yields a transverse $\Pi^{\mu\nu}$, Eq.~(\ref{final}), in every order of expansion on the intensity of the external field. One piece of this object is inherited from the form of the polarization tensor in vacuum and receives a leading correction of order $(eB)^2$, whereas the second piece is transverse in the coordinates on the graphene membrane and vanishes in the absence of the field. Direct calculation not always renders a manifestly transverse polarization operator~\cite{pol}, for instance, in ordinary QED. Spurious terms might arise as a consequence of a regularization procedure. Nevertheless, careful treatment of the regulators ensure gauge invariance is preserved for arbitrary magnetic field strength. QED$_3$ being superrenormalizable, lacks of UV-regularization issues. Nevertheless, we have presented an alternative calculation to the standard representation of the polarization tensor as a double proper time integral~\cite{Schwinger,Dittrich,schubert,Shpagin}, which manifestly preserves gauge invariance.

As an application of the vacuum polarization tensor, we have estimated the light absorption in graphene and the angle of rotation of polarization of light passing through a membrane of this material. We observe a deviation of the form $(eB)^2/\omega^4$ as compared to the vacuum result for graphene opacity. The same behavior is observed for the angle of polarization rotation. Our findings are in agreement with previously reported theoretical calculations~\cite{fial1,fial2,fial3} as well as the experimental light absorption of $2.3\%$ per graphene membrane~\cite{measure}. Further applications of the polarization tensor presented here and the effective action derived from it are under scrutiny and will be presented elsewhere.

\section*{Acknowledgments}

We acknowledge valuable discussions from Cristi\'an Villavicencio, \'Angel S\'anchez and Mar\'{\i}a Elena Tejeda. AR and SHO acknowledge CONACyT (M\'exico) for financial support for sabbatical and short visit at PUC, respectively and CIC-UMSNH under grant No. 4.22 as well as the hospitality of PUC, where the main part of this work was carried out. ML acknowledges final support from FONDECyT (Chile) grants Nos. 1130056 and 1120770. DV acknowledges support from CONICYT (Chile).

\appendix
\section*{Appendix}
\setcounter{section}{1}

In this appendix, we compute the master integral in Eq.~(\ref{master}). For this purpose, we write
\begin{equation}
I^{{\rm fg}}_{{\rm mnr}}= \int_0^1 x^{\rm f} (1-x)^{\rm g} J_{\rm mnr}(x;p)\,,\label{master2}
\end{equation}
with
\begin{equation}
J_{\rm m n r}(x;p)= \int d^3k\ \frac{k_\parallel^{\rm 2m} k_\perp^{\rm 2n}}{[k^2+p^2x(1-x)]^{\rm r}}\;.
\end{equation}
After Wick rotating to Euclidean space, writing $d^3k=\pi dk_\parallel k_\perp dk_\perp$ and with the aid of the identity
\begin{equation}
B(x,y)=2\int_0^\infty dt \ t^{2x-1}(1+t^2)^{-x-y}\;,
\end{equation}
we immediately obtain
\begin{eqnarray}
J_{\rm m n r}(x;p)&=& (-1)^{\rm m+n-r}i\pi B({\rm n}+1, {\rm r-n}-1)B\left({\rm r}+\frac{1}{2},{\rm r-m-n}-\frac{3}{2}\right)
\nonumber\\
&&\times\frac{1}{\left[p^2x(1-x) \right]^{{\rm r-m-n}-3/2}}
\;.
\end{eqnarray}
Then, the remaining integral over $x$ in Eq.~(\ref{master2}) can be performed from the definition of the beta function
\begin{equation}
B(x,y)=\int_0^1 dt \ t^{x-1}(1-t)^{y-1}\;,
\end{equation}
which finally lead us to the result~(\ref{master}).

\end{document}